# Symmetry Breaking with the SCAN Density Functional Describes Strong Correlation in the Singlet Carbon Dimer

John P. Perdew[a,b,*], Shah Tanvir ur Rahman Chowdhury[c], Chandra Shahi[a], Aaron D. Kaplan[a],

Duo Song[d], and Eric J. Bylaska[d]

[a]Department of Physics, Temple University, Philadelphia, PA 19122

[b]Department of Chemistry, Temple University, Philadelphia, PA 19122

[c]Thayer School of Engineering, Dartmouth College, Hanover, NH 03755

[d]Pacific Northwest National Laboratory, Richland, WA 99353

**Abstract:** The SCAN (strongly constrained and appropriately normed) meta-generalized gradient approximation (meta-GGA), which satisfies all 17 exact constraints that a meta-GGA can satisfy, accurately describes equilibrium bonds that are normally correlated. With symmetry breaking, it also accurately describes some *sd* equilibrium bonds that are strongly correlated. While *sp* equilibrium bonds are nearly always normally correlated, the $C_2$ singlet ground state is known to be a rare case of strong correlation in an *sp* equilibrium bond. Earlier work that calculated atomization energies of the molecular sequence $B_2$, $C_2$, $O_2$, and $F_2$ in the local spin density approximation (LSDA), the Perdew-Burke-Ernzerhof (PBE) GGA, and the SCAN meta-GGA, without symmetry breaking in the molecule, found that only SCAN was accurate enough to reveal an anomalous under-binding for $C_2$. This work shows that spin symmetry breaking in singlet $C_2$, the appearance of net up- and down-spin densities on opposite sides (not ends) of the bond, corrects that under-binding, with a small SCAN atomization-energy error more like that of the other three molecules, suggesting that symmetry-breaking with an advanced density functional might reliably describe strong correlation. This article also discusses some general aspects of symmetry breaking, and the insights into strong correlation that symmetry-breaking can bring.

**\*Corresponding author:** perdew@temple.edu, 1-215-204-1407

## Introduction

The spin-density functional theory of Kohn and Sham[1,2] is an exact in principle self-consistent-field formalism for the ground-state energy and spin densities of a system of interacting non-relativistic electrons in the presence of an external possibly-spin-dependent potential $v_\sigma(\boldsymbol{r})$. In practice, only the exchange-correlation energy as a functional of the spin densities must be approximated. The computationally-efficient approximations on the first three rungs of Jacob's Ladder[3] are single integrals of the form

$$E_{xc}^{approx}[n_\uparrow, n_\downarrow] = \int d^3 r\, n\varepsilon_{xc}^{approx}(n_\uparrow, n_\downarrow, \nabla n_\uparrow, \nabla n_\downarrow, \tau_\uparrow, \tau_\downarrow). \tag{1}$$

Here $n(\boldsymbol{r}) = n_\uparrow(\boldsymbol{r}) + n_\downarrow(\boldsymbol{r}) = \sum_{i\sigma}^{occup} |\psi_{i\sigma}(\boldsymbol{r})|^2$ is the total electron density, and

$$\tau_\sigma(r) = \sum_i^{occup} \frac{1}{2} |\nabla \psi_{i\sigma}(\boldsymbol{r})|^2 \qquad (2)$$

is the positive orbital kinetic energy density. Rung 1 of the ladder (e.g., LSDA[1,2,4]) uses only the local spin densities, rung 2 (e.g., PBE[5] GGA) adds the gradients of the spin densities, and rung 3 (e.g., SCAN[6] or r²SCAN[7] meta-GGA) adds the kinetic energy densities.

There is a nearly straight line[8] from the exact density functional theory to its first-principles approximations (e.g., LSDA, PBE, or SCAN), which proceeds from exact (if impractical) expressions for $E_{xc}[n_\uparrow, n_\downarrow]$ to the derivation of mathematical properties of the exact functional that can be satisfied by an approximation on a given rung, and finally to imposing those constraints on the approximate functional. Appropriate norms, or systems in which a given rung can be accurate for the exchange energy alone and the correlation energy alone, can be safely fitted. Appropriate norms for all three rungs include electron gases of uniform spin densities. Appropriate norms for rung 3 include closed-subshell atoms. LSDA inherits 8 exact constraints from its uniform gas norm, PBE satisfies 11 exact constraints appropriate to the GGA level, and SCAN satisfies all 17 known exact constraints that a meta-GGA can satisfy. The energies of molecules and solids, often fitted by empirical functionals, are not appropriate norms, because of an understood but imperfect and uncontrolled cancellation between errors in the approximate exchange and correlation energies. While empirical functionals interpolate between bonds, first-principles approximations predict[9] them, typically working better than empirical functionals for artificial molecules that are unlike those commonly fitted.

SCAN with a long-range van der Waals correction was more accurate than other tested functionals from the first three rungs, and more accurate than some hybrid functionals employing a fraction of exact exchange, in a 2017 test[10] on the GMTKN55 suite of 55 molecular test sets. SCAN is also accurate for insulating solids. For some strongly-correlated *sd*-bound solids, such as the cuprates[11] and manganese dioxides[12], symmetry-broken SCAN is accurate without a "Hubbard-like:"+*U* correction (sometimes interpreted as a self-interaction correction[13]). In other strongly-correlated transition-metal oxides, symmetry-broken SCAN requires a +*U* correction, but one significantly smaller than PBE requires[14]. Range-separated GGA hybrid functionals with short-range exact exchange can describe[15] many strongly-correlated solids better than LSDA or GGA and without a material-dependent parameter like *U*. Such functionals[16] and other nonlocal functionals[17] can also describe polaronic symmetry breaking.

For equilibrium bonds, a self-interaction correction seems to be much more needed for *sd* than for *sp* bonds. The Perdew-Zunger self-interaction correction[18] is first-principles, but is not reliably accurate due to the lobedness of its localized one-electron densities[19]. Perhaps in the future an improved self-interaction correction to a SCAN-like functional will lead to a reliable and widely-useful description of strong correlation via symmetry breaking. For now, any strongly-correlated equilibrium *sp* bond is of special interest as a test of the ability of symmetry breaking to describe strong correlation. Such bonds are rare, but the singlet ground state of the molecule $C_2$ is known to be strongly correlated. This molecule has an avoided crossing of two energy surfaces near its equilibrium bond length, and has been studied carefully with the full configuration interaction quantum Monte Carlo (FCIQMC) correlated-wavefunction approach.[20]

Symmetry breaking (in both wavefunction and density functional theories) and strong correlation have attracted the interest of many, notably Gustavo E. Scuseria and Richard L. Martin[15,22,23]. Strong correlation arises when degenerate or nearly degenerate Slater determinants are strongly mixed by the electron-electron pair interaction. This leads to correlation or exchange-correlation energies more negative than those that a standard approximate density functional can produce. A standard approximate functional places the total energy of a strongly-correlated state too high, and can often lower its energy by breaking a symmetry. Breaking the symmetry can break the degeneracy and return the system to a normally-correlated state whose energy the approximate functional can properly describe.

This role of symmetry breaking was recognized[24] early in the history of density functional theory: For the singlet $H_2$ molecule at its equilibrium bond length, standard approximate functionals can be accurate for the energy. As the bond length is stretched toward infinity, preserving the correct singlet (hence spin-unpolarized) ground state requires a strongly negative correlation energy which the approximate functional cannot provide. In correlated-wavefunction language, the singlet ungerade excited state is dropping toward and becoming degenerate with the singlet gerade ground state. At a critical Coulson-Fischer bond length, the approximate functional starts to produce a spin-polarized (hence symmetry-broken) self-consistent density, which is needed for a realistic $H_2$ binding energy curve. As the bond length tends to infinity, one recovers an up-spin H atom on one end of the bond and a down-spin H atom on the other, which is actually correct in the sense that such a spin-density fluctuation freezes out, not on an infinite time scale but on a macroscopic one. Thus the symmetry-breaking provides a physical picture of the strong correlation present at large bond length. In this limit, the symmetry breaking has converted a strong correlation energy into zero correlation energy (literally zero in SCAN, which is self-correlation free).

For a given Hamiltonian, eigenstates of a complete set of commuting observables (including symmetry operators that leave the Hamiltonian invariant) can always be found. In this sense, symmetry breaking is not required in exact wavefunction and density functional theories. Approximate density functionals that avoid symmetry breaking have also been constructed[25,26]. Still, as pointed out in the previous paragraph, symmetry breaking is sometimes fully real and perhaps more often revealing.

An interesting recent comment by Alex Zunger[27] suggests that all or at least many "quantum materials" (exotic extended systems, including strongly-correlated ones) can be well described by full symmetry breaking with standard approximate functionals (including in some cases $+U$-style self-interaction corrections). Much is known about symmetry breaking[28,29], but much remains to be determined, including its reliability for the description of strong correlation with advanced density functionals.

Open-shell atoms like B, C, O, and F have degenerate ground states that can be either symmetry-preserving or symmetry-breaking. LSDA, PBE, and SCAN appear not to make significant differences in energy between them. *sp* atoms do not seem to exhibit strong correlation, although bonded systems can do so. The molecules $B_2$, $C_2$, $O_2$, and $F_2$ have closed-subshell non-degenerate ground states, A recent study[30] of the atomization energies of $B_2$, $C_2$, $O_2$, and $F_2$ used symmetry-unbroken solutions for the molecules (Fig.1 of Ref. 30). The mean absolute errors for $B_2$, $O_2$, and $F_2$ were 1.7 eV in LSDA, 0.7 eV for PBE, and 0.1 eV for SCAN. The errors of LSDA and PBE were too large to permit the identification of singlet $C_2$ as the only strongly-correlated molecule in this set. But the much smaller errors of SCAN showed that singlet $C_2$ is strongly and uniquely under-bound (by 1.5 eV) in symmetry-unbroken SCAN. We will show here that this under-binding is corrected, and the error becomes similar to the errors for the other three molecules, within symmetry-broken SCAN. Thus we will show that the strong correlation in singlet $C_2$, which is far more subtle than that in highly-stretched $H_2$, can be captured by symmetry-broken SCAN.

## Computational Details

The calculations in this study were performed with the pseudopotential plane-wave NWPW[31] module implemented in the NWChem[32,33] software package. The web application EMSL Arrows[34] was used to set up and perform all the calculations. The LSDA[1,2] (in the parametrization of Ref. 4), PBE GGA[5] and SCAN meta-GGA[6] were employed to account for the exchange and correlation energy. In our plane-wave calculations, the valence electron interactions with the atomic core for carbon were approximated with generalized norm-conserving Hamann[35] pseudopotentials modified to the separable form suggested by Kleinman and Bylander[36]. The pseudopotentials used in this study were constructed with the following core radii: $r_{cs}$ = 0.800 bohr, $r_{cp}$ = 0.850 bohr and $r_{cd}$ = 0.850 bohr. The electronic wavefunctions were expanded using a plane-wave basis in a simple cubic box of L = 26 bohr with a wavefunction cutoff energy of 100 Ry and a density cutoff energy of 200 Ry.

Both restricted and unrestricted calculations with aperiodic free space boundary conditions[37,38] were performed for the singlet electronic state of $C_2$. The spin-symmetry breaking was "nudged". For the unrestricted calculations, the initial Kohn-Sham molecular orbitals were randomly generated with numerical noise to break the spin-symmetry. This is automatically done by default in NWChem.

## Results and Discussion

Table I presents our calculated atomization energies for the strongly-correlated singlet state of the carbon dimer. The results without spin-symmetry breaking in the molecule show a familiar pattern in which the atomization energy is strongly reduced from LSDA to PBE GGA to SCAN meta-GGA. The atomization energy with spin-symmetry breaking is also reduced in the same order, but much less strongly. The error for symmetry-broken SCAN is 0.0 eV, in line with the 0.1 eV mean absolute error of SCAN for the normally-correlated and hence symmetry-unbroken dimers $B_2$, $O_2$, and $F_2$ discussed in the Introduction and in Ref. 30. Molecular spin-symmetry breaking in SCAN increases the atomization energy by 1.4 eV.

The last column or "difference" in Table I is the energy lowering due to the appearance of non-zero net spin density in a singlet state that in a symmetry-preserving theory would have zero net spin density everywhere. This difference increases strongly from 0.3 eV for LSDA to 0.7 eV for PBE GGA to 1.4 eV for SCAN meta-GGA. Compare Table I of Ref. 30, in which the energy difference between spherical and non-spherical density for open-subshell B, C, O, and F atoms also increases strongly from LSDA to PBE GGA to SCAN. Added ingredients in Eq. (1) can make an approximate functional much more sensitive to the density[30], at least in some cases.

Figures 1, 2, and 3 show the calculated binding energy curves for singlet $C_2$ in LSDA, PBE GGA, and SCAN, respectively. The Coulson-Fischer point always occurs at a bond length much shorter than the equilibrium bond length (~ 1.27 Å for all three functionals, with or without symmetry breaking). The pattern of spin-symmetry breaking, as shown in these figures, is also similar for all three functionals, with up and down spin densities appearing on opposite sides (not opposite ends as in stretched $H_2$) of the bond (double in $C_2$ but single in $H_2$). The earliest spin-symmetry-broken solution in singlet $C_2$ was end-to-end[39]. A side-to-side alternation of lower energy was found in Ref. 40 with the PW91 GGA (which preceded but is numerically similar to the PBE GGA). We suspect that the symmetry breaking is revealing low-frequency spin-density fluctuations that contribute to the strong correlation in $C_2$.

The symmetry-unbroken singlet wavefunction of C$_2$ would have the total-spin quantum number $S = 0$ and thus $\langle \hat{S}^2 \rangle = S(S+1) = 0$. Symmetry breaking of the Kohn-Sham non-interacting wavefunction increases from LSDA ($\langle \hat{S}^2 \rangle = 0.66$) to PBE GGA ($\langle \hat{S}^2 \rangle = 0.89$) to SCAN meta-GGA ($\langle \hat{S}^2 \rangle = 1.07$).

The calculated Kohn-Sham orbitals explain the complicated pattern of net spin density shown in the insets of Figs. 1-3. The net bonding in C$_2$ arises from four electrons in the highest-energy occupied molecular orbitals. In the symmetry-unbroken molecule, these are two degenerate $\pi$ bonding orbitals. In the symmetry-broken molecule, these are a highest-occupied $\sigma$ orbital that is the same for spin-up and spin-down, and two next-highest-occupied banana bonds, one for spin down with its banana lobe above the bond axis and one for spin up with its banana lobe below in the page of each figure. A larger banana lobe on one side of the bond axis is accompanied by a smaller and more compact lobe of the same orbital on the other side. The net spin density has an integrated net down (blue) spin density above the bond axis and an integrated net up (orange) spin density below the bond axis, as shown in the figures. The integral of the difference in spin-up and spin-down densities is 0.353 (-0.353) from unrestricted LSDA, 0.442 (-0.442) from unrestricted PBE GGA and 0.628 (-0.628) from unrestricted SCAN meta-GGA below (above) the bond axis.

In summary, spin-symmetry breaking with the SCAN meta-GGA captures the energetics (and perhaps also the modality) of strong correlation in the singlet carbon dimer at its equilibrium geometry. For this equilibrium *sp* bond, no need for a self-interaction correction is expected. When a self-interaction correction to the SCAN-like functionals is perfected, symmetry breaking might provide a widely reliable description of strong correlation in molecules and solids.

Symmetry-broken states of finite systems may be "static or strong correlation" states that persist for a long-time (long enough to have well-defined energies), if not forever[28,29]. Access to such states is through the dynamic structure factor[24,41] or spectral function[28,29] describing the dynamic correlations that are hidden within a time-independent wavefunction, that are revealed in the expectation values of relevant time-dependent operators, and that are in fact responsible for the exchange-correlation energy that is approximated in ground-state density-functional theory.

**Acknowledgments:** JPP and STuRC were supported by the U.S. National Science Foundation under grant DMR-1939528, CMMT-Division of Materials Research, with a contribution from CTMC-Division of Chemistry. CS was supported by the U.S. Department of Energy, Office of Science, Office of Basic Energy Sciences, as part of the Computational Chemical Sciences Program, under Award No. DE-SC0018331. ADK was supported by a Temple University Presidential Fellowship. EJB and DS were supported by the U.S. Department of Energy (DOE), Office of Science, Office of Basic Energy Sciences, Chemical Sciences, Geosciences, and Biosciences (CSGB) Division through its CCS and Geosciences projects at Pacific Northwest National Laboratory (DE-AC06-76RLO-1830). We would also like to thank the National Energy Research Scientific Computing Center (NERSC), a User Facility supported by the Office of Science of the U.S. DOE under Contract No. DE-AC02-05CH11231

*Symmetry Breaking with the SCAN Density Functional Describes Strong Correlation in the Singlet Carbon Dimer*

**Table I.** Atomization energy (eV) of singlet $C_2$ at its equilibrium bond length from several nonempirical density functionals, calculated without and with spin-symmetry breaking (SB) in the molecule. All the equilibrium bond distances were obtained by DFT geometry optimization ($R_{eq}$ = 1.24 Å from restricted LSDA, 1.27 Å from unrestricted LSDA, 1.25 Å from restricted PBE GGA, 1.29 Å from unrestricted PBE GGA, 1.23 Å from restricted SCAN meta-GGA and 1.27 Å from unrestricted SCAN meta-GGA). The reference value is the FCIQMC value from Ref. 20. (1 hartree = 27.21 eV = 627.5 kcal/mol)

| Functional | without SB | with SB | difference |
|---|---|---|---|
| LSDA | 7.19 | 7.47 | 0.28 |
| PBE GGA | 6.03 | 6.75 | 0.72 |
| SCAN meta-GGA | 4.76 | 6.19 | 1.43 |
| reference |  | 6.22 |  |

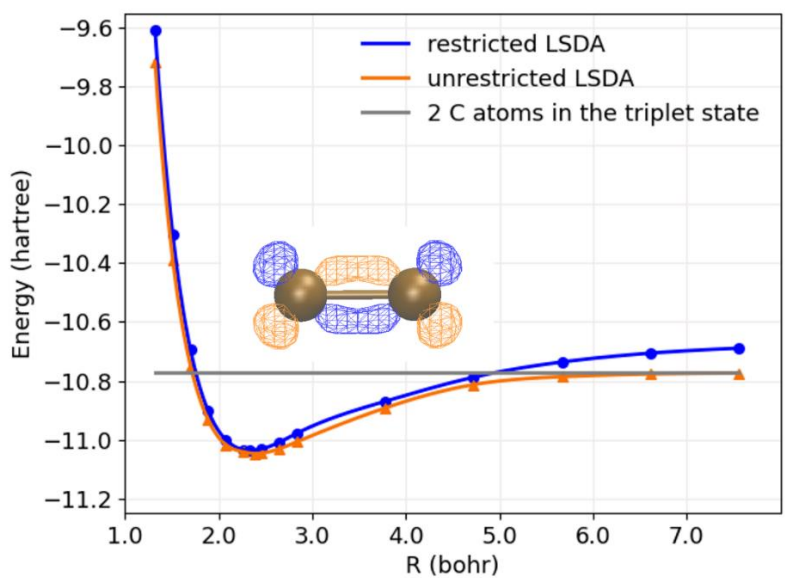

**Fig. 1.** LSDA binding energy curve for singlet $C_2$. Pseudopotential energy in hartree vs. bond length in bohr. Blue: without symmetry breaking in the molecule. Orange: with spin-symmetry breaking in the molecule. The inset shows +0.05 (orange) and -0.05 au (blue) contours surfaces of $n_\uparrow - n_\downarrow$ for the symmetry-broken solution at the equilibrium bond length $R_{eq}$ = 2.39 bohr (1.27 Å).

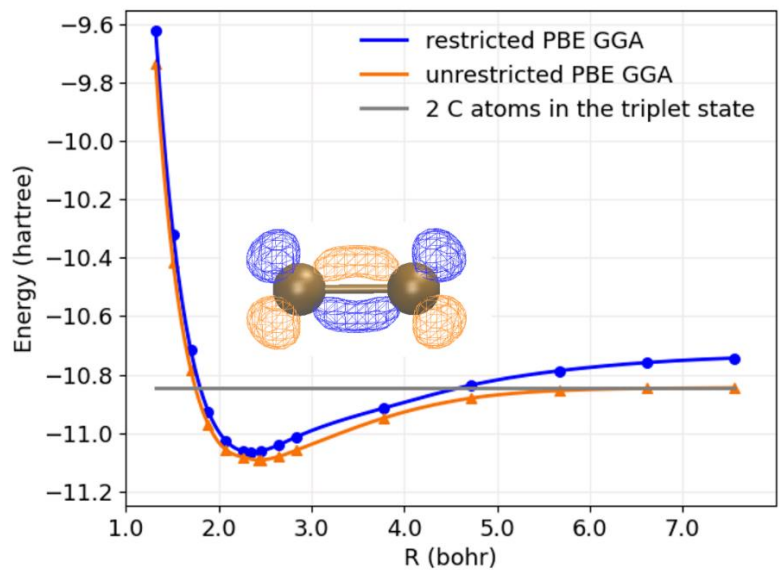

**Fig. 2.** PBE GGA binding energy curve for singlet $C_2$. Pseudopotential energy in hartree vs. bond length in bohr. Blue: without symmetry breaking in the molecule. Orange: with spin-symmetry breaking in the molecule. The inset shows +0.05 (orange) and -0.05 au (blue) contours surfaces of $n_\uparrow - n_\downarrow$ at the equilibrium bond length $R_{eq}$ = 2.43 bohr (1.29 Å).

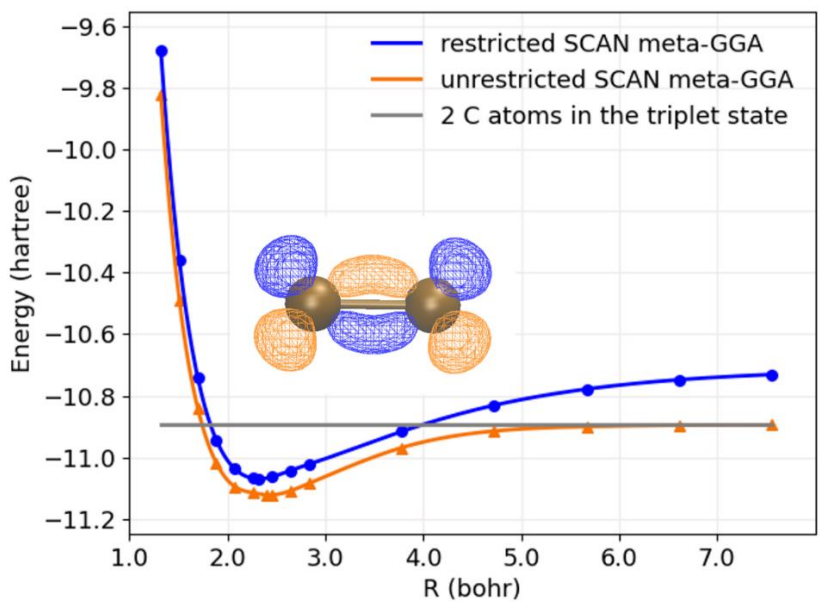

**Fig. 3.** SCAN meta-GGA binding energy curve for singlet $C_2$. Pseudopotential energy in hartree vs. bond length in bohr. Blue: without symmetry breaking in the molecule. Orange: with spin-symmetry breaking in the molecule. The inset shows +0.05 (orange) and -0.05 au (blue) contours surfaces of $n_\uparrow - n_\downarrow$ at the equilibrium bond length $R_{eq}$ = 2.41 bohr (1.27 Å).